\documentclass[twocolumn,pra,superscriptaddress,nofootinbib,longbibliography]{revtex4-2}
\usepackage{amsmath,amsfonts,amssymb,amsthm}
\usepackage{graphicx,subfigure}
\usepackage{mathrsfs}
\usepackage[mathscr]{eucal}
\usepackage{mathtools}
\usepackage[T1]{fontenc}
\usepackage{physics}
\usepackage{bm}
\usepackage[bookmarks=false]{hyperref}
\usepackage{xcolor}
\usepackage{lineno}
\usepackage{url}
\usepackage{algorithm}       
\usepackage{algpseudocode}

\hypersetup{colorlinks=true,citecolor=blue,
linkcolor=blue,urlcolor=blue,pdfstartview=FitH,
bookmarksopen=true}

\usepackage{pdfpages}
\usepackage{pgffor}
\makeatletter 
\AtBeginDocument{\let\LS@rot\@undefined} 
\makeatother 

\begin{document}
\title{Scalable protocol to coherence estimation from scarce data: Theory and experiment}
\author{Qi-Ming Ding}
\thanks{These two authors contributed equally to this work.}
\affiliation{Center on Frontiers of Computing Studies, Peking University, Beĳing 100871, China}
\affiliation{School of Computer Science, Peking University, Beĳing 100871, China}%
\affiliation{ZQuantum (Beijing) Technology Co., Ltd, Beĳing 100871, China}
\author{Ting Zhang}
\thanks{These two authors contributed equally to this work.}
\affiliation{Center on Frontiers of Computing Studies, Peking University, Beĳing 100871, China}
\affiliation{School of Computer Science, Peking University, Beĳing 100871, China}
\author{Hui Li}
\affiliation{College of Engineering and Applied Sciences, Nanjing University, Nanjing, 210093, China}
\author{Da-Jian Zhang}
\email{email: zdj@sdu.edu.cn}
\affiliation{Department of Physics, Shandong University, Jinan 250100, China}

\begin{abstract}
  Key quantum features like coherence are the fundamental resources enabling quantum advantages and ascertaining their presence in quantum systems is crucial for developing quantum technologies. This task, however, faces severe challenges in the noisy intermediate-scale quantum era. On one hand, experimental data are typically scarce, rendering full state reconstruction infeasible. On the other hand, these features are usually quantified by highly nonlinear functionals that elude efficient estimations via existing methods. In this work, we propose a scalable protocol for estimating coherence from scarce data and further experimentally demonstrate its practical utility. The key innovation here is to relax the potentially NP-hard coherence estimation problem into a computationally efficient optimization. This renders the computational cost in our protocol insensitive to the system size, in sharp contrast to the exponential growth in traditional methods. This work opens a novel route toward estimating
  coherence of large-scale quantum systems under data-scarce conditions.
\end{abstract}
\date{\today}
\maketitle

\section{Introduction}
Key quantum features like coherence, entanglement, and magic are the fundamental resources that empower quantum information processing (QIP) to outperform its classical counterpart. A vital task is therefore to ascertain the presence of these features in quantum systems, which is crucial for developing quantum technologies but faces severe challenges in the noisy intermediate-scale quantum era \cite{Preskill2018}. On one hand, due to the exponential growth of the Hilbert space with system size, known as the curse of dimensionality \cite{Tsang2020PRX,Zhang2024PRL,Zhou2025PRA}, the experimental data available are typically scarce and insufficient for full state reconstruction. This renders traditional methods like quantum state tomography (QST) infeasible. On the other hand, many resource measures are highly nonlinear, which complicates their estimation in large systems. Although state-of-the-art methods like shadow tomography have been proposed to alleviate experimental burden \cite{Elben2022NRP,Cieslinski2024PR}, the practical utility of these methods, however, has thus far been mostly confined to estimating low-degree polynomials of states \cite{Zhou2024nQI,Zhang2025PRL}, such as expectation values of observables \cite{Huang2020,Elben2020PRL}, purities \cite{Enk2012PRL,Elben2018PRL,Brydges2019S}, and statistical moments \cite{Elben2020,Neven2021nQI,Yu2021PRL,Liu2022PRL}.

These challenges are particularly pronounced when it comes to estimating quantum coherence, which represents one of the most fundamental features of quantum mechanics \cite{2017Streltsov41003,2018Hu1,Chitambar2019RoMP,Jin2021SCPM&A,Du2022SCMA,Zheng2025SCPM&A} and stands out due to its broad relevance across diverse fields, such as quantum cryptography \cite{Grosshans2002PRL,Grosshans2003N}, quantum metrology \cite{2004Giovannetti1330,2011Giovannetti222}, and quantum thermodynamics \cite{Lostaglio2015PRX,Narasimhachar2015NC,Lostaglio2015NC,Zhang2022nQI}. The underlying reason is that quantum coherence is often quantified by intricate functions that are not amenable to efficient estimation with current techniques \cite{Zhang2018PRL}. A representative instance is the relative entropy of coherence (REC), whose estimation has received  considerable attention \cite{Zhang2019PRL,Nie2019PRL,Yu2019PRA,Yuan2020nQI,Ding2021PRR} but remains intractable for large systems due to its high nonlinearity \cite{Zhang2024nQI}. Consequently, while the rapid developments in the resource theory of coherence \cite{2017Streltsov41003,2018Hu1,Chitambar2019RoMP} have underscored the pressing need for estimating coherence, significant progress has been achieved mainly for small systems so far \cite{Wang2017PRL,Zhang2018PRL,Zheng2018PRL,Ringbauer2018PRX,Zhang2019PRL,Nie2019PRL,Yu2019PRA,Yuan2020nQI,Ding2021PRR,Ma2021PRA,Zhang2021AQT,Sun2022PRA,Ray2022PRA,Li2024PRA,Zhang2024nQI}.

In this work, we propose a scalable protocol for estimating coherence from scarce data and further conduct an experiment demonstrating its practical utility. Targetting at the REC, we show that the coherence estimation problem, which is potentially NP-hard in general, can be relaxed into a computationally efficient optimization over a well-behaved objective function. This relaxation allows us to efficiently solve the problem via a fast, gradient-based algorithm. Aided by extensive numerical simulations, we show that the computational cost in our protocol, quantified by the iteration complexity, is insensitive to the system size. This is a striking feature that is beyond the reach of traditional methods, whose computational cost typically scales exponentially with the system size. We further experimentally verify the practical utility of our protocol by examining Werner states \cite{1989Werner4277}, which play a key role in various QIP tasks. We show that reliable estimates of coherence can be extracted from very limited data obtained here. This work opens up a novel route toward estimating
coherence of large-scale quantum systems under data-scarce conditions, offering a valuable tool for certifying quantum advantages in various QIP tasks.

This work is organized as follows. In Sec.~\ref{sec:preliminary}, we briefly review the significance of the REC. In Sec.~\ref{sec:theo-protocol}, we present our theoretical protocol for estimating the REC from scarce data. In Sec.~\ref{sec:num-simulation}, we numerically analyze the iteration complexity of our algorithm. In Sec.~\ref{sec:exp-dem}, we experimentally demonstrate the practical utility of our protocol by estimating the coherence of two-qubit Werner states. Finally, we conclude in Sec.~\ref{sec:conlusion} with a brief summary and outlook.

\section{Preliminaries}\label{sec:preliminary}
The resource theory of coherence \cite{2017Streltsov41003,2018Hu1,Chitambar2019RoMP} provides a rigorous framework for quantifying coherence in terms of its utility as a physical resource in QIP tasks. Within this framework, the coherence of a state $\rho$ is defined relative to a fixed reference basis, $\ket{i}$, $i=1,\cdots, d$, with $d$ denoting the Hilbert space dimension. A prominent coherence measure is the REC, defined by $C_r(\rho)=S(\rho||\rho_\textrm{diag})$, where $S$ is the quantum relative entropy and $\rho_\textrm{diag}$ is the diagonal part of $\rho$ in the chosen basis \cite{Baumgratz2014PRL}. While numerous coherence measures have been proposed
\cite{2017Streltsov41003,2018Hu1,Chitambar2019RoMP}, the REC is distinguishable for a number of reasons. First, it possesses operational significance across diverse QIP tasks. Examples include intrinsic randomness generation \cite{Yuan2015PRA}, quantum key distribution \cite{Ma2019PRA}, nonequilibrium thermodynamics \cite{Francica2019PRE}, and Bayesian metrology \cite{Lecamwasam2024PQ}. More importantly, the REC characterizes the optimal asymptotic rate in coherence distillation \cite{Winter2016PRL}, thereby endowing coherence with an operational interpretation, which is of central interest in the resource theory of coherence. Second, the REC plays a foundational role in characterizing the dynamics of coherence under noisy channels \cite{Bromley2015PRL,Yu2016PRA}. Indeed, it has been found that all measures of coherence are frozen for an initial state in a strictly incoherent channel if and only if the REC is frozen for the state \cite{Yu2016PRA}. This makes the REC of paramount importance from a practical perspective, since understanding how coherence evolves and is affected by noise is crucial for developing noise-resilient quantum technologies \cite{Zhang2019PRL}. Third, the REC provides a unifying operational bridge between coherence and other types of quantum resources.  For example, it equals the maximal relative entropy of entanglement obtainable via incoherent operations \cite{Streltsov2015PRL}, it bounds the increase of quantum discord under any incoherent channel \cite{Yao2015PRA,Ma2016PRL}, and it underlies both correlated coherence measures and coherence-based entanglement monotones \cite{Wang2017SR,Tan2018PRL}.  This line of research has provided a unified picture for understanding different quantum resources and their interplay, allowing us to gain fruitful insights into the fundamental nature of quantumness.

\section{Theoretical protocol}\label{sec:theo-protocol}

Let us consider the setting that the experimental data available consist of the expectation values $\overline{O}_i=\tr(\rho O_i)$ of $M$ observables $\{O_i\}_{i=1}^M$. We are interested in the data-scarce regime $M\ll d^2$. Apparently, the unknown state $\rho$ cannot be uniquely determined from the given data, rendering the exact value of $C_r(\rho)$ inaccessible. We are thus led to seek practically accessible estimates of $C_r(\rho)$ like upper and lower bounds. It is worth noting that lower bounds are uniquely significant for certifying whether there is enough coherence in $\rho$ for performing certain QIP tasks. In contrast, any other type of estimates, e.g., upper bounds, cannot serve this purpose, since $C_r(\rho)$ may still be small even when such estimates are large. Hence, lower bounds
are more informative than other types of estimates in practice. The most informative lower bound is \cite{Zhang2018PRL}
\begin{equation}
  \alpha=\min\left\{C_r(\rho)|\overline{O}_i=\tr(\rho O_i), i=1,\cdots, M\right\},
  \label{eq:lb}
\end{equation}
where the minimization is over all the states compatible with the data. Notably, in comparison with any other lower bounds, the bound Eq.~\eqref{eq:lb} is tightest and lies
closest to the actual value of $C_r(\rho)$. However, evaluating Eq.~(\ref{eq:lb}) is quite challenging, as entropy minimization problems are usually NP-hard and only tractable in some special cases \cite{Kovacevic2012}. To our knowledge, no efficient method has been reported for evaluating $\alpha$. In particular, our prior study \cite{Zhang2018PRL} provides a numerical method for computing $\alpha$ but is restricted to small systems, since this method requires generating exponentially many random states.

To address this challenge, our idea is to do some relaxation to render the minimization problem in Eq.~(\ref{eq:lb}) tractable.
We first impose a mild requirement on the set of observables $\{O_i\}_{i=1}^M$. We require that the basis $\{\ket{i}\}_{i=1}^d$ is included in the set $\{O_i\}_{i=1}^{M}$, that is, $O_i=\ket{i}\bra{i}$, for $i=1,\cdots, d$. This requirement is experimentally motivated because the basis is typically chosen to be the computational basis and the corresponding measurement is routinely implementable. Let $p_i=\tr(\rho\ket{i}\bra{i})$ denote the probability of observing outcome $i$ in this measurement. We denote the collection of the basis projectors and corresponding probabilities as $\bm{b}=[\ket{1}\bra{1},\cdots, \ket{d}\bra{d}]$ and $\bm{p}=[p_1,\cdots, p_d]$ so that their dot product can be written as $\bm{p}\cdot\bm{b}=\sum_{i=1}^d p_i\ket{i}\bra{i}$. Moreover, for clarity, we adopt the notations $\bm{o}=[O_{d+1},\cdots, O_{M}]$ and $\bm{q}=[\tr(\rho O_{d+1}),\cdots,\tr(\rho O_{M})]$, which are used to collectively represent the rest of the observables in the set $\{O_i\}_{i=1}^{M}$ and their associated expectation values, respectively.

\begin{figure}[t!bp]
  \centering
  \includegraphics[width=\columnwidth]{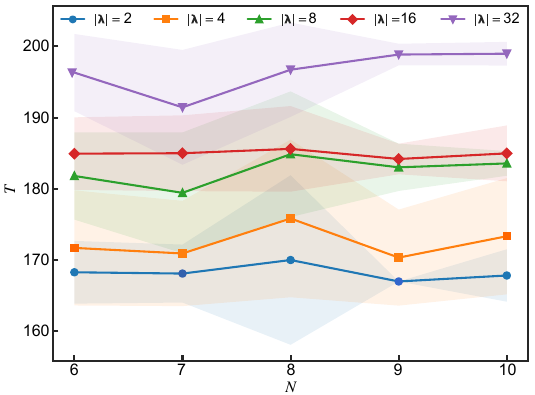}
  \caption{{Iteration number $T$ as a function of the qubit number $N$.} The values of $T$ are numerically obtained as the iteration counts required for reaching an accuracy of $10^{-5}$ with a learning rate of $0.05$. Different curves correspond to different numbers of Lagrange multipliers, $\lvert \bm{\lambda} \rvert$, indicated in the legend. Data points in the plot are obtained by averaging over $100$ random instances, with the shaded regions showing one standard deviation.}
  \label{Fig:Numeric}
\end{figure}

We next relax Eq.~\eqref{eq:lb} to render it computationally tractable. With the above notations, we can express $\alpha$ as
\begin{align}
  \alpha=\min_\rho\quad & S(\rho||\rho_\textrm{diag})\label{eq:obj} \\
  \text{s.~t.}\quad     & \tr(\rho\bm{b})=\bm{p},\quad
  \tr(\rho\bm{o})=\bm{q}.\label{eq:cons}
\end{align}
We modify the above equations to obtain a closely related optimization,
\begin{align}
  \beta=\min_\rho\quad & S(\rho||\rho_\textrm{diag})+S(\rho_\textrm{diag}||\bm{p}\cdot\bm{b})\label{eq:obj2} \\
  \text{s.~t.}\quad    &
  \tr(\rho\bm{o})=\bm{q}.\label{eq:cons2}
\end{align}
That is, we remove the constraint $\tr(\rho\bm{b})=\bm{p}$ from Eq.~(\ref{eq:cons}) while adding the term $S(\rho_\textrm{diag}||\bm{p}\cdot\bm{b})$ to Eq.~(\ref{eq:obj}). Noting that $\tr(\rho\bm{b})=\bm{p}$ if and only if $\rho_\textrm{diag}=\bm{p}\cdot\bm{b}$, we deduce that
\begin{equation}
  \alpha\geq\beta.
\end{equation}
This implies that $\beta$ is also practically significant, as it is a lower bound on the actual value of $C_r(\rho)$. Note that $S(\rho_\textrm{diag}||\bm{p}\cdot\bm{b})$ is a penalty term, which is nonnegative and vanishes if and only if the constraint $\tr(\rho\bm{b})=\bm{p}$ is fulfilled. We therefore expect that $\beta$ does not deviate too much from $\alpha$.

We now introduce an efficient procedure for numerically computing $\beta$. We show that (see Appendix \ref{app:proof} for the proof)
\begin{equation}\label{eq:main_result}
  \beta=\max_{\bm{\lambda}}\left\{-\frac{\log e}{e}\tr[2^{\log(\bm{p}\cdot \bm{b})-\bm{\lambda}\cdot \bm{o}^\prime}]-\bm{\lambda}\cdot\bm{q}^\prime\right\},
\end{equation}
that is, $\beta$ is determined through maximizing the objective function
\begin{equation}\label{eq:objective}
  f(\bm{\lambda})=-\frac{\log e}{e}\tr[2^{\log(\bm{p}\cdot \bm{b})-\bm{\lambda}\cdot \bm{o}^\prime}]-\bm{\lambda}\cdot\bm{q}^\prime.
\end{equation}
Here $\bm{o}^\prime$ is obtained by augmenting $\bm{o}$ with the identity
operator, i.e., $\bm{o}^\prime =[\bm{o}, I]$, and accordingly, $\bm{q}^\prime=[\bm{q}, 1]$. The vector $\bm{\lambda} = [\lambda_1, \dots, \lambda_{M-d+1}]$ contains the
Lagrange multipliers. Throughout this work, logarithms are taken to base 2.
The Eq.~(\ref{eq:main_result}) follows from showing that the optimization
problem in Eqs.~(\ref{eq:obj2}) and (\ref{eq:cons2}) is equivalent to the
minimum relative entropy estimation problem. The latter is known to be computationally tractable \cite{Georgiou2006IToIT,Zorzi2014ITIT}.
We highlight that $f(\bm{\lambda})$ is concave and differentiable, which allows us to employ a standard gradient-based algorithm to efficiently figure out its maximum in an iterative manner (see Appendix~\ref{sec:method} for the pseudocode of the algorithm).

Based on the above discussion, our protocol proceeds as follows: (i) measure
a set of observables $\{O_i\}_{i=1}^M$ that includes the basis
$\{\ket{i}\}_{i=1}^d$; (ii) input the resulting data into our algorithm to
iteratively compute $\beta$, which provides a reliable estimate of
$C_r(\rho)$. We clarify that how to choose the observables $\bm{o}$ is an interesting topic but is beyond the scope of this work. What we are concerned with here is to obtain an informative estimate of coherence from the given data, regardless of how $\bm{o}$ are chosen. We confirm the informativeness of $\beta$ in Appendix~\ref{sec:method2} by numerically demonstrating that $\beta$ is indeed close to $\alpha$. Notably, the number of iterations required in our algorithm, known as iteration complexity and denoted by $T$ hereafter, is a natural figure of merit for characterizing the efficiency of the algorithm. Below we demonstrate the effectiveness of our protocol through numerically analyzing $T$.

\section{Numerical simulation}\label{sec:num-simulation}

To numerically analyze $T$, we need to examine two factors that may affect $T$. One is the number of Lagrange multipliers, denoted as $\abs{\bm{\lambda}}$, which is determined by the number of data points in $\bm{q}$. This factor characterizes the dimensionality of the optimization problem. The other factor is the qubit number $N$. Figure \ref{Fig:Numeric} shows the iteration number $T$ required to reach the accuracy $10^{-5}$ with the learning rate $0.05$. Different curves correspond to different $\abs{\bm{\lambda}}$, indicated in the legend. To enhance the statistical relevance of our numerical results, we randomly select $100$ sets of observables in $\bm{o}$ and data points in $\bm{q}$ for each $\abs{\bm{\lambda}}$. The data points in the plot represent the average of the resulting values of $T$. As shown in Fig.~\ref{Fig:Numeric}, $T$ gradually increases with $\abs{\bm{\lambda}}$ for a fixed $N$, which is consistent with the intuition that a larger number of Lagrange multipliers corresponds to a more complex optimization problem. This suggests that our protocol is particularly useful in the
data-scarce regime, where $\bm{q}$ contains relatively few data points. More interestingly, $T$ exhibits little dependence on the qubit number $N$ for a fixed $\abs{\bm{\lambda}}$, indicating that the performance of our algorithm is not sensitive to the system size. This insensitivity is a key feature that supports the scalability of our protocol to large systems.

\section{Experimental demonstration}\label{sec:exp-dem}
\begin{figure}
  \centering
  \includegraphics[width=1\linewidth]{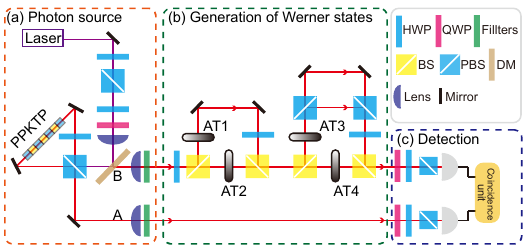}
  \caption{{Experimental setup.}~(a)~A 405 nm continuous-wave laser pumps a PPKTP crystal placed inside a Sagnac interferometer to generate polarization-entangled photon pairs via type-II spontaneous parametric down-conversion. (b)~A family of Werner states are prepared by introducing controlled decoherence into the entangled photon pairs. (c)~Polarization measurements are performed in different bases to estimate the expectation values of observables and reconstruct the quantum states via QST. DM: dichroic mirror; SPD: single-photon detector.}
  \label{fig:setup}
\end{figure}

We now apply our protocol to estimate the coherence of two-qubit Werner states \cite{1989Werner4277},
\begin{equation}\label{eq:werner}
  \rho_W(p) = p\Psi^-+ (1 - p)\frac{I}{4},
\end{equation}
where $p \in [0,1]$ and $\Psi^- = \ket{\psi^-}\bra{\psi^-}$ with $\ket{\psi^-} = (\ket{01}-\ket{10})/\sqrt{2}$. The experimental setup is shown in Fig.~\ref{fig:setup}.
We start by generating polarization-entangled photon pairs using a periodically poled potassium titanyl phosphate (PPKTP) crystal placed within a Sagnac interferometer. The interferometer is bidirectionally pumped by a 405 nm ultraviolet diode laser, as shown in Fig.~\ref{fig:setup}(a). The produced photon pairs ideally form the state $\ket{\psi^+} =(\ket{HV}+\ket{VH})/\sqrt{2}$, with horizontal ($H$) and vertical ($V$) polarizations encoding the logical states $\ket{0}$ and $\ket{1}$, respectively.

To generate $\rho_W(p)$, we take advantage of the decomposition~\cite{werner180201},
\begin{equation}
  \rho_W(p) = \frac{1+3p}{4} \Psi^-+\frac{1-p}{4} (\Psi^++\rho_{HH}+\rho_{VV}),
\end{equation}
where $\Psi^+ = \ket{\psi^+}\bra{\psi^+}$, $\rho_{HH} = \ket{HH}\bra{HH}$, and $\rho_{VV} = \ket{VV}\bra{VV}$. We generate the required mixture by applying controlled transformations to the initial state $\Psi^+$ using a combination of beam splitters (BSs), polarizing beam splitters (PBSs), half-wave plates (HWPs), and attenuators (ATs), as shown in Fig.~\ref{fig:setup}(b). The proportions of the four components are precisely adjusted by tuning the transmittance $\eta_i$ of the attenuators (ATs). Specifically, these transmittances are set according to the relations: $\eta_1=(1-p)/(2+2p)$, $\eta_2=(1+3p)/(2+2p)$, $\eta_3=(1-p)/2$, and $\eta_4=(1+p)/2$. We experimentally generate the desired Werner states, $\rho_W(p)$, at eight distinct values of $p \in \{0, 0.1, 0.2, 1/3, 0.4, 0.6, 0.8, 1\}$, and perform QST on each to obtain an average fidelity of $0.9681\pm 0.0117$. More details about the experimental setup and the state preparation can be found in Appendix \ref{app:exp-detail}.

Immediately after preparing the Werner states, we proceed with the measurement procedure, as depicted in Fig.~\ref{fig:setup}(c). The measurement module consists of a quarter-wave plate (QWP), an HWP, and a PBS. By precisely rotating the HWP and QWP to certain prescribed angles,
we can implement the projective measurements in various polarization bases.
In particular, this allows us to measure the expectation values of some key observables such as $XX$, $ZZ$, and $YY$ from the observed coincidence counts.

\begin{figure}[t!bp]
  \includegraphics[width=1\columnwidth]{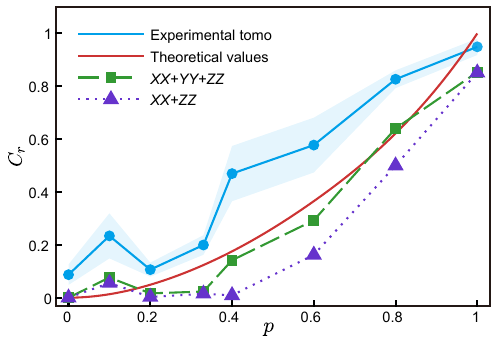}
  \caption{{Coherence estimation for Werner states.} The blue solid line with a shaded error region shows coherence estimates from QST, with error bars denoting the standard deviation from 1000 repetitions. The red solid line is the theoretical coherence for the ideal state. The purple dotted line marks the lower bound $\beta$ on coherence estimated using the  expectation values of observables $ZZ$ and $XX$, while the green dashed line represents $\beta$ obtained with the expectation values of observables $ZZ$, $XX$, and $YY$. Increasing the number of observables notably improves the tightness of the lower bound.}
  \label{Fig:est}
\end{figure}

To estimate the coherence of $\rho_W(p)$ using our protocol, we perform the
projective measurement of the observable $ZZ$. This allows us to efficiently extract
the probability distribution in the computational basis. We consider two
choices for $\bm{o}$. In the first, $\bm{o}$ consists solely of the observable $XX$. The
coherence estimates for this choice, produced from our numerical algorithm, are represented by the purple dotted
line in Fig.~\ref{Fig:est}. The second choice is to let $\bm{o}$ comprise the two observables $XX$ and $YY$. This choice
yields coherence estimates represented by the green dashed line. Also, for comparison,
Fig.~\ref{Fig:est} presents the theoretical coherence of the ideal state
(red solid line) and the coherence obtained directly from the experimentally
prepared states (blue solid line). The shaded region around the blue line
denotes the standard deviation resulted from 1000 experimental repetitions with
Poissonian statistics.

It is worth noting from Fig.~\ref{Fig:est} that, even with minimal experimental effort---measuring only the observable $XX$ in conjunction with $ZZ$---our protocol can yield a reasonably tight lower bound for the coherence (see the purple dotted line). This result highlights the practical significance of our protocol for inferring quantum properties with limited measurement resources. The gap between the estimated bound and the
tomographic results arises naturally, since the available data do not uniquely
specify the underlying state. Increasing the number of measured observables
generally tightens the bound. This point can be seen by comparing the purple dotted line with the green dashed line in Fig.~\ref{Fig:est}. The latter, obtained from measuring $XX$, $YY$, and $ZZ$, markedly improves the coherence estimation and visibly narrows the gap between the estimated lower bound and the ideal theoretical coherence (red solid line). This systematic convergence towards the theoretical coherence with
additional experimental input underscores the effectiveness of our protocol.

\section{Summary and outlook}\label{sec:conlusion}

We have developed and experimentally demonstrated a scalable, non-tomographic protocol for efficiently estimating coherence of unknown states from scarce data. The key theoretical advance here is to relax the entropy minimization problem defined by
Eq.~(\ref{eq:lb}) into a closely related, computationally tractable optimization.
This reformulation allows us to efficiently compute a tight lower bound on coherence using a fast, gradient-based algorithm.

Numerical simulation shows that the performance of our algorithm, characterized by the iteration complexity, is largely insensitive to the system size. This indicates the scalability of our protocol to large systems. Our photonic experiment on two-qubit Werner states, on the other hand, illustrates the practical utility of our protocol, demonstrating that it can yield reliable coherence estimates even with limited measurement data. The present work therefore opens the possibility of efficiently estimating coherence for large-scale systems under data-scarce conditions.

While we focus on the REC in this work, the methodology adopted here may be generalized to estimating other physical properties. For example, the REC is closely related to entanglement \cite{Streltsov2015PRL,Yao2015PRA,Ma2016PRL,Wang2017SR,Tan2018PRL}, which is another quantum resource of central interest in quantum information science. It would be interesting to explore whether our protocol can be adapted to estimate entanglement from scarce data. Another intriguing direction is to extend our protocol to estimating coherence in multipartite systems \cite{Radhakrishnan2016PRL}, where the complexity of the optimization problem may increase significantly. We leave these directions for future work.

\section*{Acknowledgements}
We are grateful for the beneficial discussions with Dr. Xiao Yuan. This work is supported by the Beijing Natural Science Foundation (Grant No.~1254053), the National Natural Science Foundation of China (Grant No.~12275155), and the Shandong Provincial Young Scientists Fund (Grant No.~ZR2025QB16). The scientific calculations in this paper have been done on the HPC Cloud Platform of Shandong University as well as the High-performance Computing Platform of Peking University.

\section*{Data Availability}
The data that support the findings of this work are openly available \cite{github}.

\appendix

\section{Proof of Eq.~(\ref{eq:main_result})}\label{app:proof}

Let us recall the optimization problem:
\begin{align}
  \beta=\min_\rho\quad & S(\rho||\rho_\textrm{diag})+S(\rho_\textrm{diag}||\bm{p}\cdot\bm{b})\label{eq:obj-method} \\
  \text{s.~t.}\quad    &
  \tr(\rho\bm{o})=\bm{q},\label{eq:cons1-method}\\
  & \rho\geq 0,\quad\tr\rho=1.\label{eq:cons2-method}
\end{align}
Here we have explicitly included the constraints $\rho\geq 0$ and $\tr\rho=1$ to ensure that $\rho$ is a density operator. Noting that 
\begin{equation}
    \tr\left[\rho\log(\rho_\textrm{diag})\right]=\tr\left[\rho_\textrm{diag}\log(\rho_\textrm{diag})\right],
\end{equation}
we have
\begin{equation}\label{eq:entropy_diff_diag}
    S(\rho||\rho_\textrm{diag})=\tr\left[\rho\log(\rho)\right]-\tr\left[\rho_\textrm{diag}\log(\rho_\textrm{diag})\right].
\end{equation}
Likewise, the equality
\begin{equation}
    \tr\left[\rho_\textrm{diag}\log(\bm{p}\cdot\bm{b})\right]=\tr\left[\rho\log(\bm{p}\cdot\bm{b})\right]
\end{equation}
yields
\begin{equation}\label{eq:entropy_diff_p_b}
    S(\rho_\textrm{diag}||\bm{p}\cdot\bm{b})=\tr\left[\rho_\textrm{diag}\log(\rho_\textrm{diag})\right]-\tr\left[\rho\log(\bm{p}\cdot\bm{b})\right].
\end{equation}
Using Eqs.~(\ref{eq:entropy_diff_diag}) and (\ref{eq:entropy_diff_p_b}), we can simplify the objective (\ref{eq:obj-method}) as 
\begin{equation}\label{eq:obj-simplified}
    S(\rho||\rho_\textrm{diag})+S(\rho_\textrm{diag}||\bm{p}\cdot\bm{b})=S(\rho||\bm{p}\cdot\bm{b}).
\end{equation}
Besides, letting $\bm{o}^\prime=[\bm{o},I]$ and $\bm{q}^\prime=[\bm{q},1]$, we can remove the constraint $\tr\rho=1$ in Eq.~(\ref{eq:cons2-method}) so that the optimization can be taken over the cone of positive semidefinite operators. Using this result and Eq.~(\ref{eq:obj-simplified}), we can reformulate the optimization problem in Eqs.~(\ref{eq:obj-method}), (\ref{eq:cons1-method}), and (\ref{eq:cons2-method}) as
\begin{align}
  \beta=\min_\rho\quad & S(\rho||\bm{p}\cdot\bm{b})\label{eq:obj-new} \\
  \text{s.~t.}\quad    &
  \tr(\rho\bm{o}^\prime)=\bm{q}^\prime,\label{eq:cons1-new}\\
  & \rho\geq 0.\label{eq:cons2-new}
\end{align}
This form is equivalent to the minimum relative entropy estimation studied in Refs.~\cite{Georgiou2006IToIT,Zorzi2014ITIT}, which is known to be tractable. We can solve the problem by introducing the Lagrangian
\begin{equation}\label{eq:lagrangian}
    \mathcal{L}(\rho,\bm{\lambda})=S(\rho||\bm{p}\cdot\bm{b})+\bm{\lambda}\cdot \left[ \tr(\rho\bm{o}^\prime)-\bm{q}^\prime \right],
\end{equation}
where $\bm{\lambda} = [\lambda_1, \dots, \lambda_{M-d+1}]$ is known as the vector of
Lagrange multipliers. Following Ref.~\cite{Zhang2018PRL}, we can compute $\beta$ as
\begin{equation}\label{eq:beta-minmax}
    \beta = \max_{\bm{\lambda}} \min_{\rho\geq 0} \mathcal{L}(\rho,\bm{\lambda}).
\end{equation}
It is shown \cite{Georgiou2006IToIT,Zorzi2014ITIT} that the optimal $\rho^*$ attaining the minimum is 
\begin{equation}\label{eq:optimal-rho}
    \rho^*=2^{\log(\bm{p}\cdot\bm{b})-\log(e)I-\bm{\lambda}\cdot\bm{o}^\prime}.
\end{equation}
Inserting Eq.~(\ref{eq:optimal-rho}) into Eq.~(\ref{eq:lagrangian}), we obtain, after some algebra, 
\begin{equation}\label{eq:lagrangian-min}
    \min_{\rho\geq 0} \mathcal{L}(\rho,\bm{\lambda}) = -\frac{\log(e)}{e}\tr\left[2^{\log(\bm{p}\cdot\bm{b})-\bm{\lambda}\cdot\bm{o}^\prime}\right]-\bm{\lambda}\cdot\bm{q}^\prime.
\end{equation}
Substituting this equation into Eq.~(\ref{eq:beta-minmax}), we arrive at
\begin{equation}
    \beta = \max_{\bm{\lambda}} \left\{ -\frac{\log(e)}{e}\tr\left[2^{\log(\bm{p}\cdot\bm{b})-\bm{\lambda}\cdot\bm{o}^\prime}\right]-\bm{\lambda}\cdot\bm{q}^\prime \right\}.
\end{equation}
Lastly, it should be pointed out that the function $f(\bm{\lambda})$ introduced in the main text is $\min_{\rho\geq 0} \mathcal{L}(\rho,\bm{\lambda})$, that is, $f(\bm{\lambda})=\min_{\rho\geq 0} \mathcal{L}(\rho,\bm{\lambda})$. 

\section{Gradient-based algorithm}\label{sec:method}


We see that $f(\bm{\lambda})$ is concave, since it is the so-called Lagrange dual function \cite{Boyd2004}. This property ensures that $f(\bm{\lambda})$ admits only global maxima, thereby circumventing the issue of local maxima. Furthermore, we deduce from Eq.~(\ref{eq:objective}) that
$f(\bm{\lambda})$ is differentiable over the parameter domain. This allows us to figure out its maximum via standard gradient-based algorithms. Specifically, using the differentiation formula for exponential operators \cite{Wilcox1967JoMP}, we can compute the gradient of $f(\bm{\lambda})$, denoted by $\nabla f(\bm{\lambda})$, as follows:
\begin{equation}\label{eq:gradient}
  \nabla f(\bm{\lambda}) = \frac{1}{e}\tr\left[2^{\log(\bm{p}\cdot\bm{b})-\bm{\lambda}\cdot\bm{o}^\prime}\bm{o}^\prime\right]-\bm{q}^\prime.
\end{equation}
Then, iteratively updating the parameters $\bm{\lambda}$ using gradient ascent, we can converge to the maximum of $f(\bm{\lambda})$. The pseudocode of the algorithm is presented in Algorithm~\ref{alg:gradient_descent}.

\begin{algorithm}[H] 
  \caption{Gradient Ascent for Computing $\beta$}
  \label{alg:gradient_descent}
  \begin{algorithmic}[1]
    \State \textbf{Input:} experimental data $\bm{p}$, $\bm{q}$, learning rate $\eta > 0$, tolerance $\epsilon > 0$;
    \State \textbf{Initialize:} $\lambda_0\leftarrow$ a random guess; $k \leftarrow 0$;
    \Repeat
    \State \text{Compute the gradient $\nabla f(\bm{\lambda}_k)$ according to Eq.~(\ref{eq:gradient})};
    \State \text{Update the parameters: $\bm{\lambda}_{k+1} \leftarrow \bm{\lambda}_k + \eta \nabla f(\bm{\lambda}_k)$};
    \Until{$\|\nabla f(\bm{\lambda}_{k+1})\| \le \epsilon$};
    \State \textbf{Output:} $f(\bm{\lambda}_{k+1})$.
  \end{algorithmic}
\end{algorithm}

\section{Verifying the closeness of $\beta$ to $\alpha$}\label{sec:method2}

Here we numerically demonstrate that $\beta$ is indeed close to $\alpha$.
Note that no scalable method is currently available to us for efficiently computing $\alpha$. We therefore examine the case of two qubits, for which the numerical method from our prior work \cite{Zhang2018PRL} is capable of providing accurate estimates for $\alpha$. As a representative instance, we set $\bm{o}$ to only consist of the observable $XX$. Then, using the method from our prior work \cite{Zhang2018PRL} and the algorithm presented in the preceding subsection, we can
carry out numerical simulations to compute both $\alpha$ and $\beta$ in this setting.

\begin{figure}[!t]
  \centering
  \includegraphics[width=1\columnwidth]{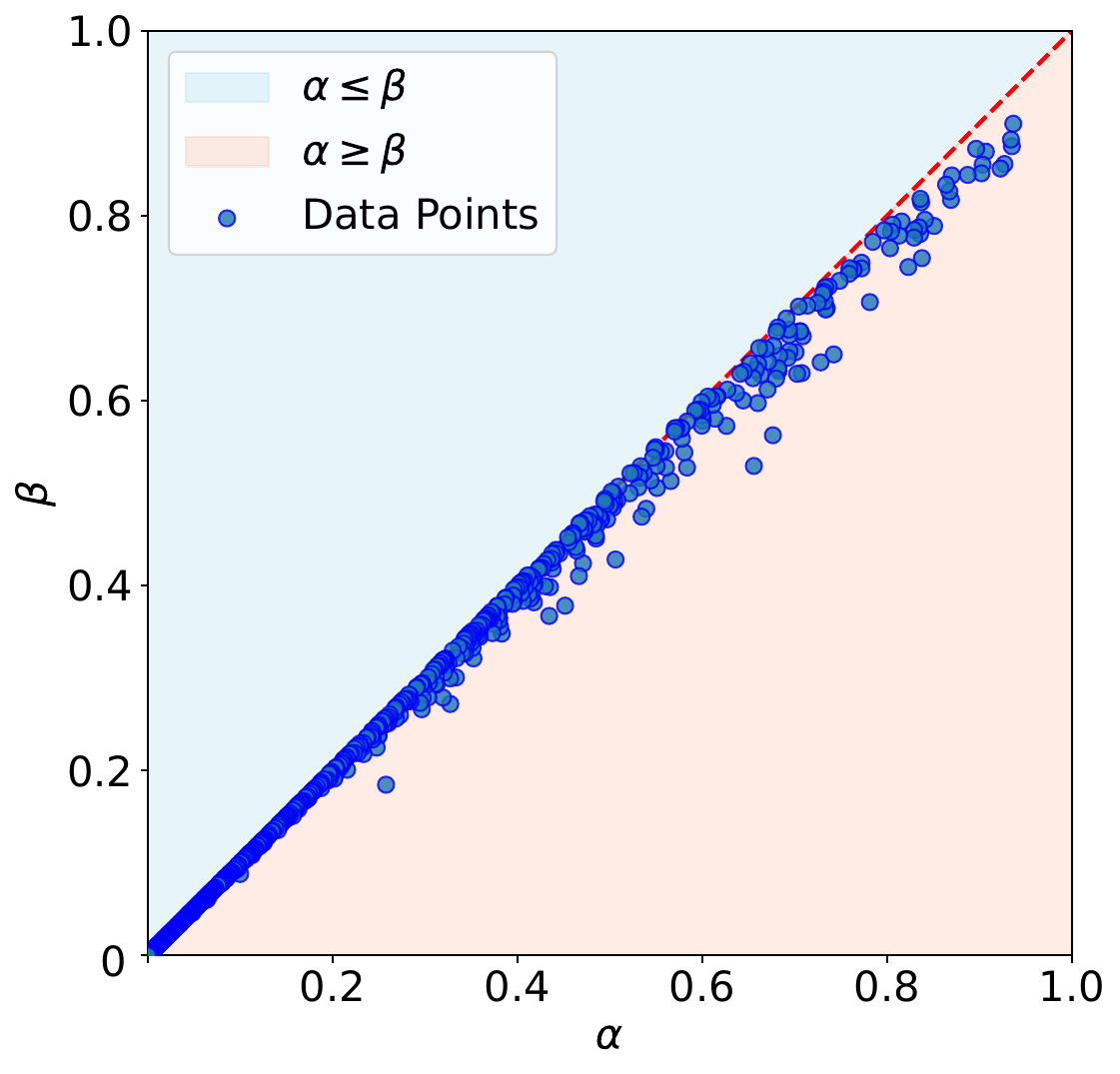}
  \caption{{Numerical results illustrating the closeness of $\beta$ to $\alpha$.} Each data point represents a pair $(\alpha,\beta)$, with $\alpha$ on the horizontal axis and $\beta$ on the vertical axis. The values of $\alpha$ and $\beta$ are obtained from the numerical method in our prior work \cite{Zhang2018PRL} and the algorithm presented in the preceding section, respectively. A total of $500$ data points are produced by randomly choosing the values of $\bm{p}$ and $\bm{q}$. The region is partitioned into two parts respectively with $\alpha\leq\beta$ and  $\alpha\geq\beta$, as indicated in the figure legend, separated by the dashed red line.}
  \label{Fig:sm-fig1}
\end{figure}

The results are displayed in Fig.~\ref{Fig:sm-fig1}. Each pair $(\alpha,\beta)$ is represented by a point in this figure, with $\alpha$ on the horizontal axis and $\beta$ on the vertical axis. By randomly choosing the values of $\bm{p}$ and $\bm{q}$, we generate $500$ data points in total. To clear see the physical meaning of these data points, we divide the region in the plot into two parts: one corresponding to $\alpha\leq\beta$ and the other to $\alpha\geq\beta$, as indicated in the figure legend. The two parts are separated by the dashed red line.
It is clear that all the data points are located on the half region with $\alpha\geq\beta$. This confirms the practical significance of $\beta$ as a lower bound on $\alpha$. More importantly, we see that all the data points cluster tightly around the dashed red line, demonstrating that $\beta$ is indeed an accurate approximation of $\alpha$.

In addition, to exclude the possibility that the behavior observed in Fig.~\ref{Fig:sm-fig1} is merely coincidental and unique to the above setting, we have conducted extensive numerical experiments across various different settings and choices of observables. These experiments consistently yield similar results and reinforce the robustness of our findings.

\section{Experimental details}\label{app:exp-detail}
In our experiment, several optical components play essential roles in manipulating and analyzing photonic qubits. Polarization control is primarily realized by half-wave plates (HWPs) and quarter-wave plates (QWPs), whose optical axes are rotated by angles $\theta$ and $\zeta$ with respect to the vertical polarization, respectively. The HWP implements the unitary transformation
\begin{equation}
U_{\mathrm{HWP}}(\theta) =
\begin{pmatrix}
-\cos 2\theta & -\sin 2\theta \\
-\sin 2\theta & \cos 2\theta
\end{pmatrix},
\end{equation}
while the QWP corresponds to
\begin{equation}
U_{\mathrm{QWP}}(\zeta) = \tfrac{1}{\sqrt{2}}
\begin{pmatrix}
1+i\cos 2\zeta & i\sin 2\zeta \\
i\sin 2\zeta & 1-i\cos 2\zeta
\end{pmatrix}.
\end{equation}
Polarizing beam splitters (PBSs) are employed as polarization-dependent routers that transmit horizontal polarization and reflect vertical polarization. In addition, non-polarizing 50:50 beam splitters (BSs) are used to divide incoming photons into transmitted and reflected paths with equal probability, independent of polarization. These components together form the basis of our optical platform for implementing and characterizing polarization-encoded qubits.

\subsection{Polarization-entangled photon source}

\begin{figure}[!t]
    \centering
\includegraphics[width=\linewidth]{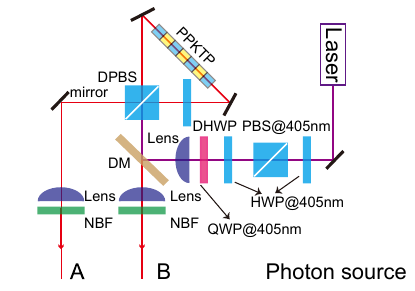}
    \caption{Schematic illustration of the setup for generating polarization-entangled photon pair.}
    \label{fig:source}
\end{figure}

The source of polarization-entangled photon pairs, shown in Fig.~\ref{fig:source}, is a Sagnac interferometer containing a $15$-mm-long periodically poled KTP crystal with a $10.025~\mu\mathrm{m}$ poling period, bidirectionally pumped by a continuous-wave $405~\mathrm{nm}$ laser. A half-wave plate at $22.5^\circ$ prepares the pump in the state $\ket{+}_P=(\ket{H}_P+\ket{V}_P)/\sqrt{2}$, which is focused to a waist of about $38~\mu\mathrm{m}$ at the crystal center. Inside the loop, the horizontally polarized pump travels clockwise and generates photon pairs in the $\ket{HV}$ state via type-II SPDC. The vertically polarized component is rotated to horizontal by a $45^\circ$ HWP, propagates counterclockwise, and likewise produces $\ket{HV}$. These two indistinguishable amplitudes are superposed at the output PBS, yielding the maximally entangled Bell state $\ket{\Psi^+}=(\ket{HV}+\ket{VH})/\sqrt{2}$. The interferometer is compactly arranged within an area of roughly $600~\mathrm{mm}\times450~\mathrm{mm}$.

\begin{table*}[t!bp]
\centering
\begin{tabular}{c|c|c}
\hline
Target state & Optical operation & Probability amplitude \\
\hline
$\ket{\psi^-}$ & HWP at $90^\circ$ + BS network & $\tfrac{\eta_{2} \eta_{4}}{16}$ \\
$\ket{\psi^+}$ & Attenuation via AT1 + BS network & $\tfrac{\eta_{1} \eta_{4}}{16}$ \\
$\ket{HH}$ & Conversion from $\ket{\psi^\pm}$ via AT3 & $\tfrac{(\eta_{1}+\eta_{2}) \eta_{3}}{32}$ \\
$\ket{VV}$ & Conversion from $\ket{\psi^\pm}$ via AT3 & $\tfrac{(\eta_{1}+\eta_{2}) \eta_{3}}{32}$ \\
\hline
\end{tabular}
\caption{Conversion of the initial $\ket{\psi^+}$ state into the components of the Werner state. $\eta_{i}$ denote the transmittances of attenuators AT$i$.}
\label{tab:Werner}
\end{table*}

The detected brightness of the source is $9042 \pm 1252$ pairs/s/mW, and the entangled state fidelity reaches $0.9917 \pm 0.0012$ with respect to the ideal Bell state, as quantified by the fidelity
$
F(\rho,\sigma)=\left(\mathrm{Tr}\sqrt{\sqrt{\rho}\,\sigma\,\sqrt{\rho}}\right)^2,
$
averaged over 15 consecutive daily measurements.
\subsection{Generation of Werner state}
The Werner state can be expressed as a probabilistic mixture of four pure components~\cite{werner180201},
\begin{widetext}
\begin{equation}
\rho_{W}(p) = \frac{1+3p}{4}\ket{\psi^-}\!\bra{\psi^-} 
+ \frac{1-p}{4}\ket{\psi^+}\!\bra{\psi^+} 
+ \frac{1-p}{4}\left( \ket{HH}\!\bra{HH} + \ket{VV}\!\bra{VV} \right).
\label{eq:Werner}
\end{equation}
\end{widetext}
To reproduce this distribution experimentally, we start from the entangled state $\ket{\psi^+}$ and probabilistically convert it into the desired ensemble of $\{\ket{\psi^-},\ket{\psi^+},\ket{HH},\ket{VV}\}$ by means of half-wave plates (HWPs), beam splitters (BSs), and adjustable attenuators (ATs). The transformation processes and their associated probabilities are summarized in Table~\ref{tab:Werner}.

\begin{figure*}[]
    \centering
    \includegraphics[width=\textwidth]{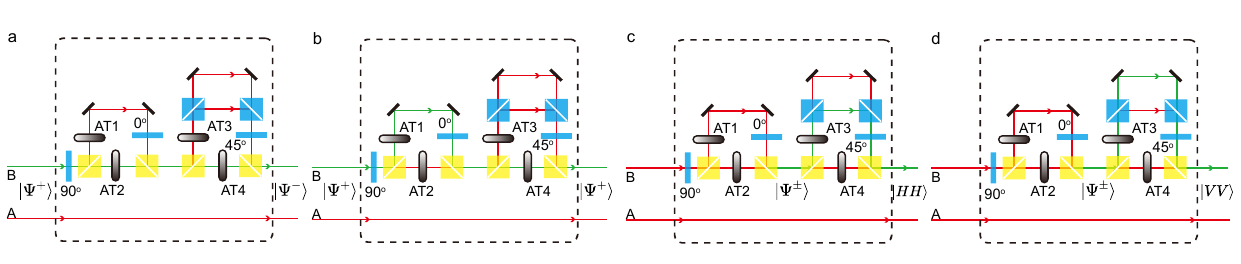}
    \caption{Schematic illustration of the setup for generating Werner states.}
    \label{fig:wer}
\end{figure*}

By incoherently mixing the four outputs on a beam splitter, the normalized state takes the form
\begin{widetext}
\begin{equation}
\rho_{\mathrm{exp}}(p) = 
\frac{\eta_{2} \eta_{4}}{N}\ket{\psi^-}\!\bra{\psi^-} +
\frac{\eta_{1} \eta_{4}}{N}\ket{\psi^+}\!\bra{\psi^+} +
\frac{(\eta_{1}+\eta_{2})\eta_{3}}{2N}\left( \ket{HH}\!\bra{HH} + \ket{VV}\!\bra{VV} \right),
\label{eq:rho_exp}
\end{equation}
\end{widetext}
where the normalization factor is
$N = (\eta_{1}+\eta_{2})(\eta_{3}+\eta_{4})$. Comparing Eq.~(\ref{eq:rho_exp}) with the definition of the Werner state in Eq.~(\ref{eq:Werner}), one obtains the required transmittance settings:
\begin{equation}
\eta_{1} = \frac{1-p}{2+2p},\quad
\eta_{2} = \frac{1+3p}{2+2p},\quad
\eta_{3} = \frac{1-p}{2},\quad
\eta_{4} = \frac{1+p}{2}.
\label{eq:transmittance}
\end{equation}

\bibliography{refs}



\end{document}